\RequirePackage{lineno} 
\documentclass[prd,twocolumn,showpacs,groupedaddress,superscriptaddress,amsmath,amssymb]{revtex4}

\usepackage{graphicx}
\usepackage{dcolumn}
\usepackage{bm}
\usepackage{amssymb}
\usepackage{enumerate}
\usepackage{lineno}

\newcommand\ainv{A'\to invisible}

\newcommand\g{\gamma}
\newcommand\ma{m_{A'}}

\newcommand\Na{{N}_{A'}}
\newcommand\ea{e^- Z \to e^- Z A'; A' \to invisible}
\newcommand\emu{e^- Z \to e^- Z \g; \g \to \mu^+ \mu^-}

\def\address{\@ifstar{\address@star}%
  {\@ifnextchar[{\address@optarg}{\address@noptarg}}}

\begin{document}

\title{Search for invisible decays of sub-GeV dark photons in missing-energy events at the CERN SPS}
%
\affiliation{\it Universit\"at Bonn, Helmholtz-Institut f\"ur Strahlen-und Kernphysik, 53115 Bonn, Germany} 
\affiliation{\it  Joint Institute for Nuclear Research, 141980 Dubna, Russia}
\affiliation{\it CERN, European Organization for Nuclear Research, CH-1211 Geneva, Switzerland}
\affiliation{\it Institute for Nuclear Research, 117312 Moscow, Russia}
\affiliation{\it P.N. Lebedev Physics Institute, Moscow, Russia, 119 991 Moscow, Russia}
\affiliation{\it Skobeltsyn Institute of Nuclear Physics, Lomonosov Moscow State University, Moscow, Russia}
\affiliation{\it Physics Department, University of Patras, Patras, Greece} 
\affiliation{\it State Scientific Center of the Russian Federation Institute for High Energy Physics of National Research Center 'Kurchatov Institute' (IHEP), 142281 Protvino, Russia}
\affiliation{\it Tomsk Polytechnic University, 634050 Tomsk, Russia}
\affiliation{\it Universidad T\'{e}cnica Federico Santa Mar\'{i}a, 2390123 Valpara\'{i}so, Chile}
\affiliation{\it ETH Z\"urich, Institute for Particle Physics, CH-8093 Z\"urich, Switzerland}

\author{D.~Banerjee}\affiliation{\it ETH Z\"urich, Institute for Particle Physics, CH-8093 Z\"urich, Switzerland}
\author{V.~Burtsev}\affiliation{\it Tomsk Polytechnic University, 634050 Tomsk, Russia}
\author{D.~Cooke}\affiliation{\it ETH Z\"urich, Institute for Particle Physics, CH-8093 Z\"urich, Switzerland}
\author{P.~Crivelli}\affiliation{\it ETH Z\"urich, Institute for Particle Physics, CH-8093 Z\"urich, Switzerland}
\author{E.~Depero}\affiliation{\it ETH Z\"urich, Institute for Particle Physics, CH-8093 Z\"urich, Switzerland}
\author{A.~V.~Dermenev}\affiliation{\it Institute for Nuclear Research, 117312 Moscow, Russia}
\author{S.~V.~Donskov}\affiliation{\it State Scientific Center of the Russian Federation Institute for High Energy Physics of National Research Center 'Kurchatov Institute' (IHEP), 142281 Protvino, Russia}
\author{F.~Dubinin}\affiliation{\it P.N. Lebedev Physics Institute, Moscow, Russia, 119 991 Moscow, Russia}
\author{R.~R.~Dusaev}\affiliation{\it Tomsk Polytechnic University, 634050 Tomsk, Russia}
\author{S.~Emmenegger}\affiliation{\it ETH Z\"urich, Institute for Particle Physics, CH-8093 Z\"urich, Switzerland}
\author{A.~Fabich}\affiliation{\it CERN, European Organization for Nuclear Research, CH-1211 Geneva, Switzerland}
\author{V.~N.~Frolov}\affiliation{\it  Joint Institute for Nuclear Research, 141980 Dubna, Russia}
\author{A.~Gardikiotis}\affiliation{\it Physics Department, University of Patras, Patras, Greece}
\author{S.~N.~Gninenko\footnote{Corresponding author, Sergei.Gninenko@cern.ch}
}\affiliation{\it Institute for Nuclear Research, 117312 Moscow, Russia}
\author{M.~H\"osgen}\affiliation{\it Universit\"at Bonn, Helmholtz-Institut f\"ur Strahlen-und Kernphysik, 53115 Bonn, Germany}
\author{V.~A.~Kachanov}\affiliation{\it State Scientific Center of the Russian Federation Institute for High Energy Physics of National Research Center 'Kurchatov Institute' (IHEP), 142281 Protvino, Russia}
\author{A.~E.~Karneyeu}\affiliation{\it Institute for Nuclear Research, 117312 Moscow, Russia}
\author{B.~Ketzer}\affiliation{\it Universit\"at Bonn, Helmholtz-Institut f\"ur Strahlen-und Kernphysik, 53115 Bonn, Germany}
\author{D.~V.~Kirpichnikov}\affiliation{\it Institute for Nuclear Research, 117312 Moscow, Russia}
\author{M.~M.~Kirsanov}\affiliation{\it Institute for Nuclear Research, 117312 Moscow, Russia}
\author{I.~V.~Konorov}\affiliation{\it P.N. Lebedev Physics Institute, Moscow, Russia, 119 991 Moscow, Russia} 
\author{S.~G.~Kovalenko}\affiliation{\it Universidad T\'{e}cnica Federico Santa Mar\'{i}a, 2390123 Valpara\'{i}so, Chile}
\author{V.~A.~Kramarenko}\affiliation{\it Skobeltsyn Institute of Nuclear Physics, Lomonosov Moscow State University, Moscow, Russia}
\author{L.~V.~Kravchuk}\affiliation{\it Institute for Nuclear Research, 117312 Moscow, Russia}
\author{ N.~V.~Krasnikov}\affiliation{\it Institute for Nuclear Research, 117312 Moscow, Russia}
\author{S.~V.~Kuleshov}\affiliation{\it Universidad T\'{e}cnica Federico Santa Mar\'{i}a, 2390123 Valpara\'{i}so, Chile}
\author{V.~E.~Lyubovitskij}\affiliation{\it Tomsk Polytechnic University, 634050 Tomsk, Russia}
\author{V.~Lysan}\affiliation{\it  Joint Institute for Nuclear Research, 141980 Dubna, Russia}
\author{V.~A.~Matveev}\affiliation{\it  Joint Institute for Nuclear Research, 141980 Dubna, Russia}
\author{Yu.~V.~Mikhailov}\affiliation{\it State Scientific Center of the Russian Federation Institute for High Energy Physics of National Research Center 'Kurchatov Institute' (IHEP), 142281 Protvino, Russia}
\author{V.~V.~Myalkovskiy}\affiliation{\it  Joint Institute for Nuclear Research, 141980 Dubna, Russia}
\author{V.~D.~Peshekhonov\footnote{Deceased}}\affiliation{\it  Joint Institute for Nuclear Research, 141980 Dubna, Russia}
\author{D.~V.~Peshekhonov}\affiliation{\it  Joint Institute for Nuclear Research, 141980 Dubna, Russia}
\author{O.~Petuhov}\affiliation{\it Institute for Nuclear Research, 117312 Moscow, Russia} 
\author{V.~A.~Polyakov}\affiliation{\it State Scientific Center of the Russian Federation Institute for High Energy Physics of National Research Center 'Kurchatov Institute' (IHEP), 142281 Protvino, Russia}
\author{B.~Radics}\affiliation{\it ETH Z\"urich, Institute for Particle Physics, CH-8093 Z\"urich, Switzerland}
\author{A.~Rubbia}\affiliation{\it ETH Z\"urich, Institute for Particle Physics, CH-8093 Z\"urich, Switzerland}
\author{V.~D.~Samoylenko}\affiliation{\it State Scientific Center of the Russian Federation Institute for High Energy Physics of National Research Center 'Kurchatov Institute' (IHEP), 142281 Protvino, Russia}
\author{V.~O.~Tikhomirov}\affiliation{\it P.N. Lebedev Physics Institute, Moscow, Russia, 119 991 Moscow, Russia}
\author{D.~A.~Tlisov}\affiliation{\it Institute for Nuclear Research, 117312 Moscow, Russia} 
\author{A.~N.~Toropin}\affiliation{\it Institute for Nuclear Research, 117312 Moscow, Russia}
\author{A.~Yu.~Trifonov}\affiliation{\it Tomsk Polytechnic University, 634050 Tomsk, Russia}
\author{B.~Vasilishin}\affiliation{\it Tomsk Polytechnic University, 634050 Tomsk, Russia}
\author{G.~Vasquez Arenas}\affiliation{\it Universidad T\'{e}cnica Federico Santa Mar\'{i}a, 2390123 Valpara\'{i}so, Chile}
\author{P.~Ulloa}\affiliation{\it Universidad T\'{e}cnica Federico Santa Mar\'{i}a, 2390123 Valpara\'{i}so, Chile}
\author{K.~Zhukov}\affiliation{\it P.N. Lebedev Physics Institute, Moscow, Russia, 119 991 Moscow, Russia}
\author{K.~Zioutas}\affiliation{\it Physics Department, University of Patras, Patras, Greece} 

%
%
\collaboration{The NA64 Collaboration\footnote{https://na64.web.cern.ch}}\noaffiliation
\vskip 0.25cm

\date{\today}

\begin{abstract}
  We report on a direct search for sub-GeV dark photons ($A'$) which  might   be produced in the reaction  $e^- Z \to e^- Z A'$ via kinetic mixing with  photons by 100 GeV electrons incident on an active target in the NA64 experiment at the CERN SPS. The $A'$s would  decay invisibly into dark matter particles resulting in events with large missing energy. No evidence for such decays was found  with $2.75\cdot 10^{9}$ electrons on target. We  set new limits on the $\gamma-A'$ mixing strength  and exclude the invisible $A'$ with a mass   $\lesssim 100$ MeV as an explanation of the muon $g_\mu-2$ anomaly.  
  
\end{abstract}
\maketitle

Despite the intensive searches at the LHC and in non-accelerator experiments Dark Matter (DM) still is  a great 
puzzle.   Though stringent constraints obtained on DM coupling to Standard Model (SM) particles   ruled out many DM models, little is known about the origin and dynamics of the  dark sector itself. 
One difficulty so far is that DM   can be probed only through its gravitational interaction.  
  An exciting possibility is that in addition to gravity,  a new force between the dark sector and visible matter  transmitted by a new vector  boson $A'$ (dark photon) might  exist. Such $A'$ could have 
a mass $m_{A'}\lesssim 1$ GeV - associated with a spontaneously broken gauged $U(1)_D$ symmetry- and couple to 
the SM  through kinetic mixing  with the ordinary photon, $-\frac{1}{2}\epsilon F_{\mu\nu}A'^{\mu\nu}$ ,  parameterized by the mixing strength  $\epsilon \ll 1$ \cite{Okun:1982xi, Galison:1983pa, Holdom:1985ag}.
This has motivated a worldwide theoretical and experimental effort  towards  dark forces and other portals between the visible and dark sectors, see \cite{jr, report} for a review. 
\begin{figure*}[tbh!]
\includegraphics[width=0.9\textwidth]{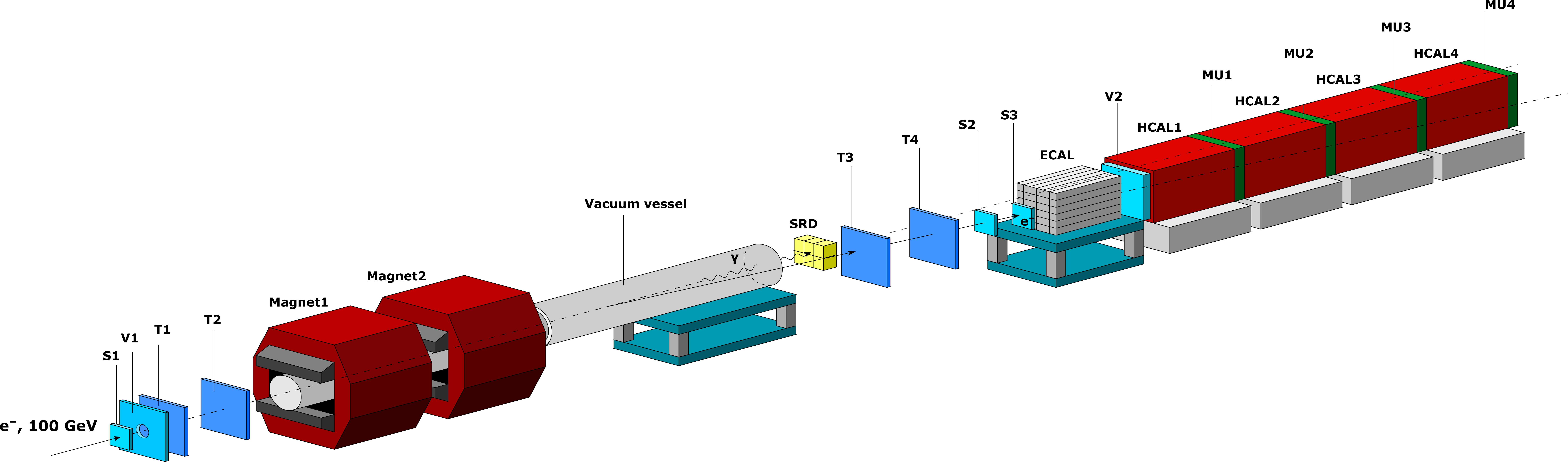}%
\caption{Schematic illustration of the setup to search for $\ainv$  decays of the bremsstrahlung $A'$s 
produced in the reaction  $eZ\rightarrow eZ A'$ of 100 GeV e$^-$ incident  on the active ECAL target.}
 \label{setup}
\end{figure*} 
 An additional motivation has been provided by hints on astrophysical signals of dark matter \cite{nima}, as well as the 3.6 $\sigma$ deviation from the SM prediction of the muon anomalous magnetic moment $g_\mu-2$ \cite{g-2},  which can be explained by a sub-GeV  $A'$ with the coupling $\epsilon\simeq 10^{-3}$ \cite{gk, fayet, mp}. Such small values of $\epsilon$ could naturally be obtained  from loop effects  of particles charged under both the dark and SM  $U(1)$ interactions with  a typical 1-loop value  $\epsilon = e g_D/16\pi^2$ \cite{Holdom:1985ag}, where $g_D$ is the coupling constant of the $U(1)_D$ gauge interactions. Various theoretical and phenomenological aspects of light vector bosons  very weakly coupled to  quarks and leptons
 have been also studied in pioneer papers by Fayet \cite{Fayet}.

  \par If the $A'$ is the lightest state in the dark sector,  then it would decay mainly visibly, i.e., typically to SM leptons $l = e, \mu$ or hadrons, which could be used to detect it. Previous beam dump \cite{jdb} -\cite{sarah1}, fixed target \cite{apex,merkel,merkel1}, collider \cite{babar, curt,  babar1}, and rare meson decay \cite{bern}- \cite{kloe3} experiments have already put stringent  constrains on  the mass $m_{A'}$	and $\epsilon$ of such dark photons excluding,  in particular,  
  the parameter region favored by the $g_\mu-2$ anomaly.
\par However, in the presence of light dark states,  in particular dark matter, with the masses $<\ma$, the $A'$  would predominantly  decay invisibly  into those particles provided that  $g_D >\epsilon e$. Models introducing such invisible  $A'$ offer new intriguing possibilities to explain the $g_\mu-2$ and various other anomalies \cite{Lee:2014tba} and are subject to different experimental   constraints  \cite{ei,psmilli,hd, Aubert:2008as}.  The most severe limits on the invisible sub-GeV $A'$s decays have been obtained from the results of beam dump experiments LSND \cite{lsnd} and E137 \cite{e137th},  under assumptions on the strength of the coupling $g_D$,  and properties of the DM decay particles.  
In this Letter we report the first results from the experiment NA64   specifically designed for a direct  search of  the $\ainv$ decay at the CERN SPS.  
\par The method of the search is as follows \cite{Gninenko:2013rka, Andreas:2013lya}. If the $A'$ exists it could be produced via the kinetic mixing with bremsstrahlung photons  in the reaction
 of high-energy electrons scattering off  nuclei of an active target  of a hermetic detector, followed by the prompt $\ainv$  decay into dark matter particles ($\chi$):
\begin{equation}
\ea
\label{ea}
\end{equation}  
  A fraction $f$ of the  primary beam energy  $E_{A'} = f E_0$ is carried away by $\chi$'s which penetrate the 
detector without interactions  resulting in an event with zero-energy deposition. While the remaining part $E_e=(1-f)E_0$ is deposited in the target  by the  scattered electron. 
Thus, the  occurrence of $A'$  produced in the reaction \eqref{ea} would appear as an excess of events whose signature is  a single e-m shower in the target with energy $E_e$ accompanied by a significant missing energy $E_{miss}=E_{A'}= E_0 - E_e$  above those expected from backgrounds. Here we assume that the $\chi$s have to traverse the detector without  decaying visibly in order to give a missing energy signature. No any other assumptions on the nature of the $\ainv$ decay are made.  
\par  The  NA64 detector  is schematically shown in Fig.~\ref{setup}.
The experiment  employed  the upgraded 100 GeV  electron beam from the H4 beamline.  
The  beam  has a maximal intensity $\simeq (3-4)\cdot 10^6$ per  SPS spill of 4.8 s produced by the primary 450 GeV/c proton beam  with an intensity of few 10$^{12}$ protons on target.  The detector utilized 
the beam defining  scintillator (Sc)  counters S1-S3, and  magnetic spectrometer  consisting of two successive  dipole magnets with the integral  magnetic field of $\simeq$7 T$\cdot$m  and a low-material-budget tracker. The tracker was a set of two upstream Micromegas chambers (T1, T2) and two downstream GEM stations (T3, T4)   allowing the measurements of $e^-$ momenta with the precision $\delta p/p \simeq 1\%$ \cite{Banerjee:2015eno}. The magnets also served as an effective filter rejecting low energy component of the beam. 
To enhance the electron identification  the synchrotron radiation (SR) emitted by electrons 
 was used for their efficient  tagging. A 15 m long vacuum vessel between the magnets and the ECAL was installed to minimize absorption of the SR
photons detected immediately at the downstream end of the vessel with a SR detector (SRD), which was either
an array of BGO crystals  or a PbSc sandwich calorimeter of a very fine segmentation \cite{Gninenko:2013rka}. By using the SRD the initial level  of the hadron contamination in the  beam $\pi/e^- \lesssim 10^{-2}$ was further suppressed by a factor $\simeq 10^3$.   
The detector was also equipped with an active target, which is an  electromagnetic (e-m) calorimeter (ECAL) for  measurement of the the electron energy with the accuracy $\delta E/E \simeq 10\%/\sqrt{E}$. 
The ECAL is a  matrix of $6\times 6 $  Shashlik-type modules  
 assembled from  Pb and Sc plates with wave-shifting fiber read-out. Each module is $\simeq 40$ radiation 
 lengths.   
 Downstream the ECAL  the detector is equipped with a high-efficiency veto counter V2, and a massive, hermetic hadronic calorimeter (HCAL)
 of $\simeq 30$ nuclear interaction lengths. The HCAL served  as a dump to completely absorb  and measure  the energy  of hadronic secondaries produced in the $e^- A \to anything$ interactions  in the target. 
Four muon plane counters, MU1-MU4,  located between the HCAL  modules were used for the muon identification in the final state. 
\begin{figure*}[tbh!!]
\includegraphics[width=0.5\textwidth]{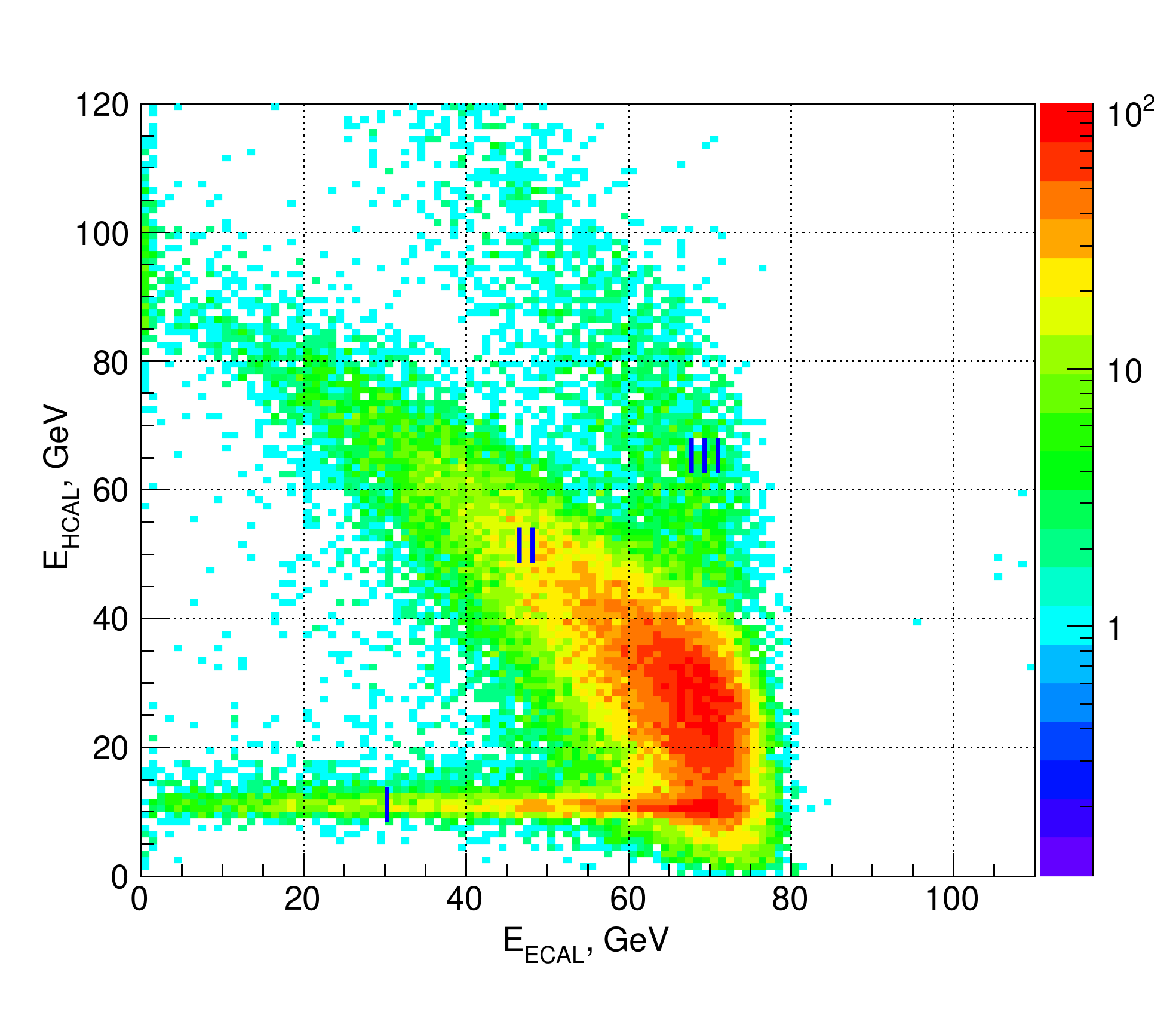}
\hspace{-0.cm}{\includegraphics[width=0.5\textwidth]{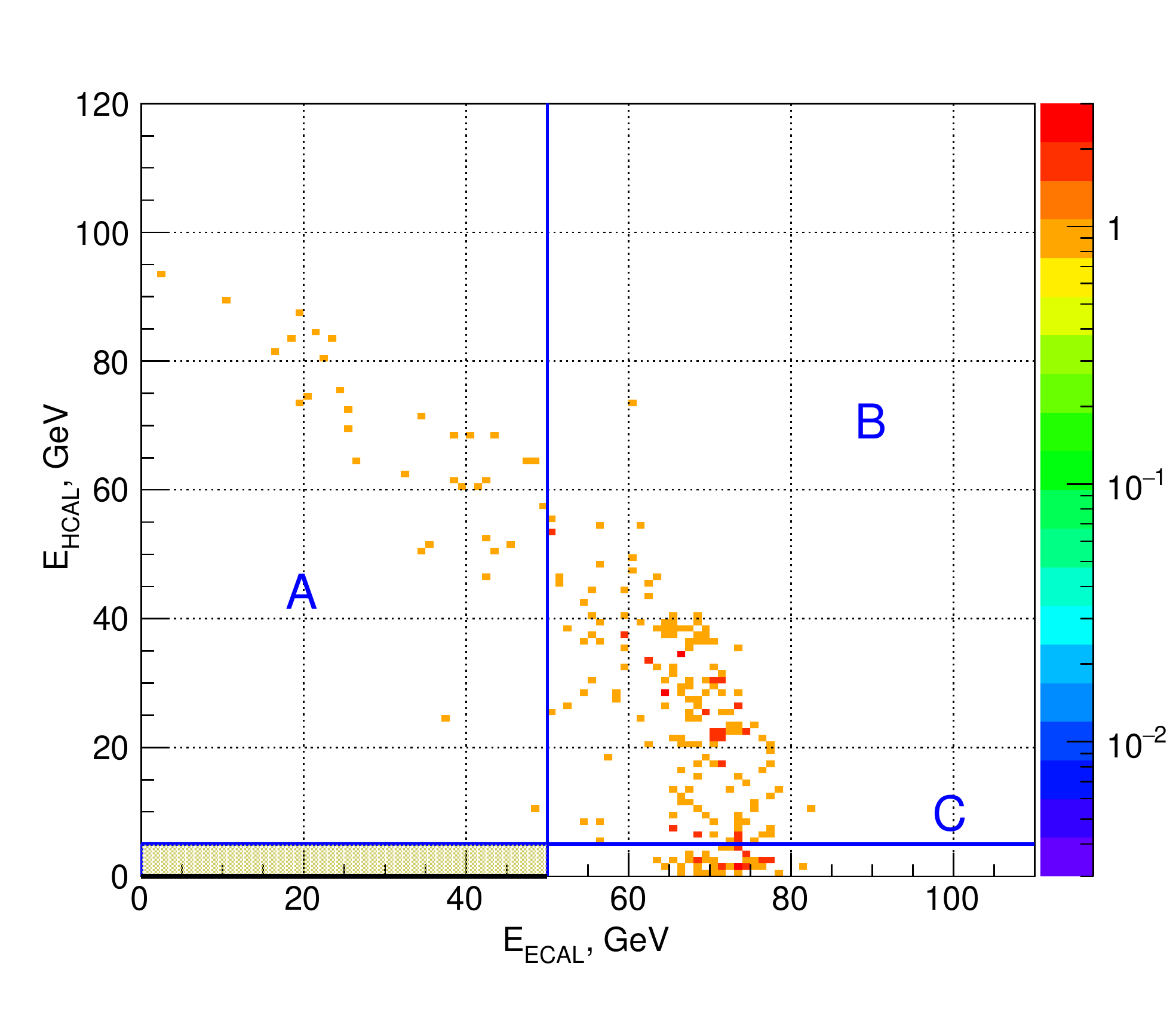}}
\caption{The left panel shows the measured  distribution of events in the ($E_{ECAL}$;$E_{HCAL}$) plane from the combined BGO and PbSc run data at the earlier phase of the analysis. Another plot shows the same distribution  after applying all selection criteria. The dashed area is the signal box  region which is open. The side bands A and C are the one  used for the background estimate inside the signal box. For illustration purposes 
the size of the signal box along $E_{HCAL}$-axis is increased by a factor five. }
\label{ecvshc}
\end{figure*}   
 The events were collected with the hardware trigger  requiring  an in-time cluster in the ECAL with the energy $E_{ECAL} \lesssim 80$ GeV. 
 The results reported here came mostly from a set of data in which $n_{eot}=1.88\cdot 10^9$ of electrons on target (eot) were collected with the beam intensity  $\simeq 1.4\cdot 10^6$ e$^-$ per spill
with the PbSc calorimeter. While a smaller sample of $n_{eot}=0.87\cdot 10^9$ and an  intensity  $I_e=0.3\cdot 10^6$ e$^-$ was also recorded with the BGO detector.  Data of these two runs (hereafter called the BGO and PbSc run) were analyzed with similar selection criteria and finally  summed up, taking into account the corresponding normalization factors.
\par In order to avoid biases in the determination of selection criteria for candidate events, a blind analysis was performed. Candidate events are expected to have the missing energy in the range $50 < E_{miss} < 100 $ GeV, 
which was defined by taking into account the energy spectrum of $A'$s emitted in the primary reaction 
\eqref{ea} by $e^\pm$ from the e-m shower generated by the beam $e^-$s in the ECAL target \cite{gkkk}. Events  from a signal box ($E_{ECAL} < 50~{\text GeV }; E_{HCAL} < 1~{\text GeV }$) 
 were excluded from the analysis of the data until the validity of the background estimate in this region was established. For  the selection criteria optimization, 10\% of the data was used, while the full sample was used for the	background estimate. The	number	of	signal	candidate events  were counted	after unblinding.
A detailed Geant4 based Monte Carlo (MC) simulation  was used to study the detector performance and acceptance, to simulate background sources, and to select cuts and estimate the reconstruction efficiency.
\par The left panel in Fig.~\ref{ecvshc} shows the distribution of the events from the reaction 
$e^- Z \to anything$ in the  $(E_{ECAL}; E_{HCAL})$ plane measured with  $2.75\cdot 10^9$ eot. Here, $E_{HCAL}$ is the sum of the energy deposited in the first two HCAL modules. About $5\cdot 10^4$ events were  selected with the loose cut requiring in-time energy deposition in the SRD within the SR range emitted by $e^-$. Events from the area I in Fig.~\ref{ecvshc} originate from the rare QED dimuon production,  dominated by the 
reaction   $\emu$, of the muon pair photoproduction by  a hard bremsstrahlung photon conversion on a target nuclei and   characterized by  the energy of $ \simeq 10$ GeV deposited by the dimuon pair in the HCAL. This  process    was used as a benchmark  allowing  to verify the reliability of the MC simulation and  estimate the systematic uncertainties in the signal reconstruction efficiency in the energy range predicted by simulations. The same  selection cuts were  applied to both signal and reference channel, in order to cross-check  systematic uncertainties. The dimuon production  was also used as a reference for the background prediction. 
The region II shows  the SM events from the hadron electroproduction in the target which satisfy 
the energy conservation $E_{ECAL} + E_{HCAL} \simeq 100$ GeV  within the energy resolution of the detectors. 
The leak of these events to the signal box due to the energy resolution is was found to be negligible. 
The events from the region III whose fraction is  a few $10^{-2}$ are mostly due to pile-up of $e^-$ and 
beam hadrons. 
\par  The candidate events were selected with the criteria chosen to maximize the acceptance for MC signal events  and to minimize the numbers of background events, respectively. The following  selection criteria were applied:
  i) The incoming particle track  should have a small angle w.r.t. the beam axis  to reject large angle tracks from the upstream $e^-$ interactions. 
ii)  The energy deposited in the SRD detector should be within the SR range emitted by $e^-$s and in-time with the trigger;
iii)  The lateral and longitudinal shape of the shower in the ECAL should be  consistent with the  one expected for the signal shower \cite{gkkk};
iv) No activity  in V2. 
Only  $\simeq 300 $ events passed these criteria from combined 
BGO and PbSc runs.
\begin{table}[tbh!] 
\begin{center}
\caption{Expected numbers of events in the signal box  from different background sources estimated for 
$2.75\cdot 10^{9}$ eot.}\label{tab:table2}
\vspace{0.15cm}
\begin{tabular}{lr}
\hline
Source of background& Events\\
\hline
loss of e$^-$ energy due to  punchthrough $\g$s& $<0.001$\\
loss of hadrons from $e^-Z\to e^- + hadrons$ & $<0.01$\\
loss or $\mu \to e \nu \nu $ decays & \\
of muons from $\emu$&$<0.01$\\
$e^-$ interactions in the beamline materials&$0.03$\\
$\mu\to e \nu \nu$, $\pi,K \to e \nu$, $K_{e3}$ decays  & $ 0.03$ \\
pile-up of low energy e$^-$ and $\mu,\pi, K$ & \\ 
followed by their decays & $0.05$ \\
$\mu, \pi, K$ interactions in the target & $0.02$ \\
\hline 
Total    &          0.15  \\
\hline
\end{tabular}
\end{center}
\end{table}
\par   The  search for the $\ainv$ decays  requires particular attention to backgrounds.
Every process  with a track and a single  e-m cluster in the ECAL  was considered as a potential source of background. There are several sources which may fake the $\ainv$ signal, e.g. upstream $e^-$ interactions,
$\mu \to e \nu \nu,~š\pi,~ K \to e \nu,~ K_{e3}$ decays in-flight,  energy leakage from particle punch-through in the HCAL, processes due to pile-up of two or more particles, and  instrumental effects due to energy loss through cracks in the upstream detector coverage. The selection cuts  to eliminate these backgrounds have been chosen such that they do not affect the shape of the true $E_{miss}$ spectrum. \\
\par Two independent methods were used for the background estimation in the signal region. The first method is based on the MC.  Due to the small coupling strength of the $A'$  reaction \eqref{ea} occurs typically 
with a rate $\lesssim 10^{-9}$ per incoming electron. To  study the SM distribution and  background  at this level is very time-consuming. Consequently, we have evaluated with MC  all known backgrounds to the extent that it is possible. Events from particle interactions or decays in the beam line, pile-up activity created from them,  hadron punch-trough  from the target and the HCAL were  included in the simulation of all background events.  Small event-number  backgrounds such as the decays of the beam  $\mu, \pi, K$ or $\mu$ from the reaction of dimuon production were  simulated with the full statistics of the data.     Large event-number processes, e.g.	
upstream beam interactions,  punch-through  of secondary hadrons were  also	studied  extensively, although simulated samples with statistics similar to the data were not feasible. 
To eliminate  possible instrumental effects not present in the MC, the uniformity scan of the  central part of the ECAL target was performed with $e^-$ by using T3 and T4. We also examined the number of events observed in several regions around the signal box, which  were statistically consistent with the estimates. 
\par Two largest sources of background are expected from the beam $\mu,~\pi,~K$ decays in-flight. In one case, when, e.g. a pion  passes through the  vacuum   vessel it could  knock electrons off the downstream window, which hit the SRD creating a fake tag for a 100 GeV $e^-$. Then the pion could decay into $e\nu$ in the upstream 
ECAL region thus producing the fake signal. Similar background is caused by  the
pile-up of an electron  from the low-energy beam tail ($\lesssim 60-80$ GeV)  and a beam $\mu,\pi$, or $K$. The electron could emit the amount of SR energy above the threshold which is detected in the SRD as a tag of  100 GeV $e^-$ and then is deflected by the magnets out of the detector's acceptance angle. While the accompanied muon or hadron could then  decay in flight. For both sources  the dominant  background came from the $K_{e3}$ decays.  
The mistakenly tagged $\mu$,  and $\pi$ and $K$  could also interact in the target producing an e-m like cluster below 50 GeV though the $\mu Z \to \mu Z \g$ or $\pi, K$ charge-exchange reactions in the target, 
accompanied by the poorly detected scattered $\mu$, or secondary hadrons, respectively.
Another background is due to $e^-$ interactions with the beamline  materials resulting in $e^-$ energy loss.   Table I summarizes the  conservatively estimated number of background events inside the signal box. 
 The expected number of background events is $0.15 \pm 0.03 (stat) \pm 0.06 (syst)$. The systematic error includes the uncertainties in the amount of passive material 
for upstream $e^-$ interactions,  and in the  cross sections of the of $\pi,K$ charge-exchange reactions  on lead (30\%).   
\par The second method used the background estimate extracted from the data themselves.
 MC signal events  and the background extrapolated from sidebands A and C shown in the right panel of Fig. \ref{ecvshc} were used. Events in the region A are pure neutral hadronic secondaries produced 
 by electrons in the ECAL target, while events from the region C are likely  from the  $e^-$ 
 interactions in the downstream part of the beamline accompanied by  bremsstrahlung photons absorbed in the HCAL.   The yield of the background events was estimated by extrapolating the observed events to the signal region assessing the systematic uncertainties by varying the background fit models. Using this 
   we  obtained a second background estimate of 0.4 $\pm$ 0.3 events.  
    The  background estimates with the two methods are in agreement with each other within errors.  After determining all the selection criteria and estimating background levels, we examined the events in the signal box and found no candidates, as shown  in  Fig.~\ref{ecvshc}. The conclusion  that the background is small is confirmed by the data. 
\begin{figure}[tbh]
\begin{center}
\includegraphics[width=0.5\textwidth]{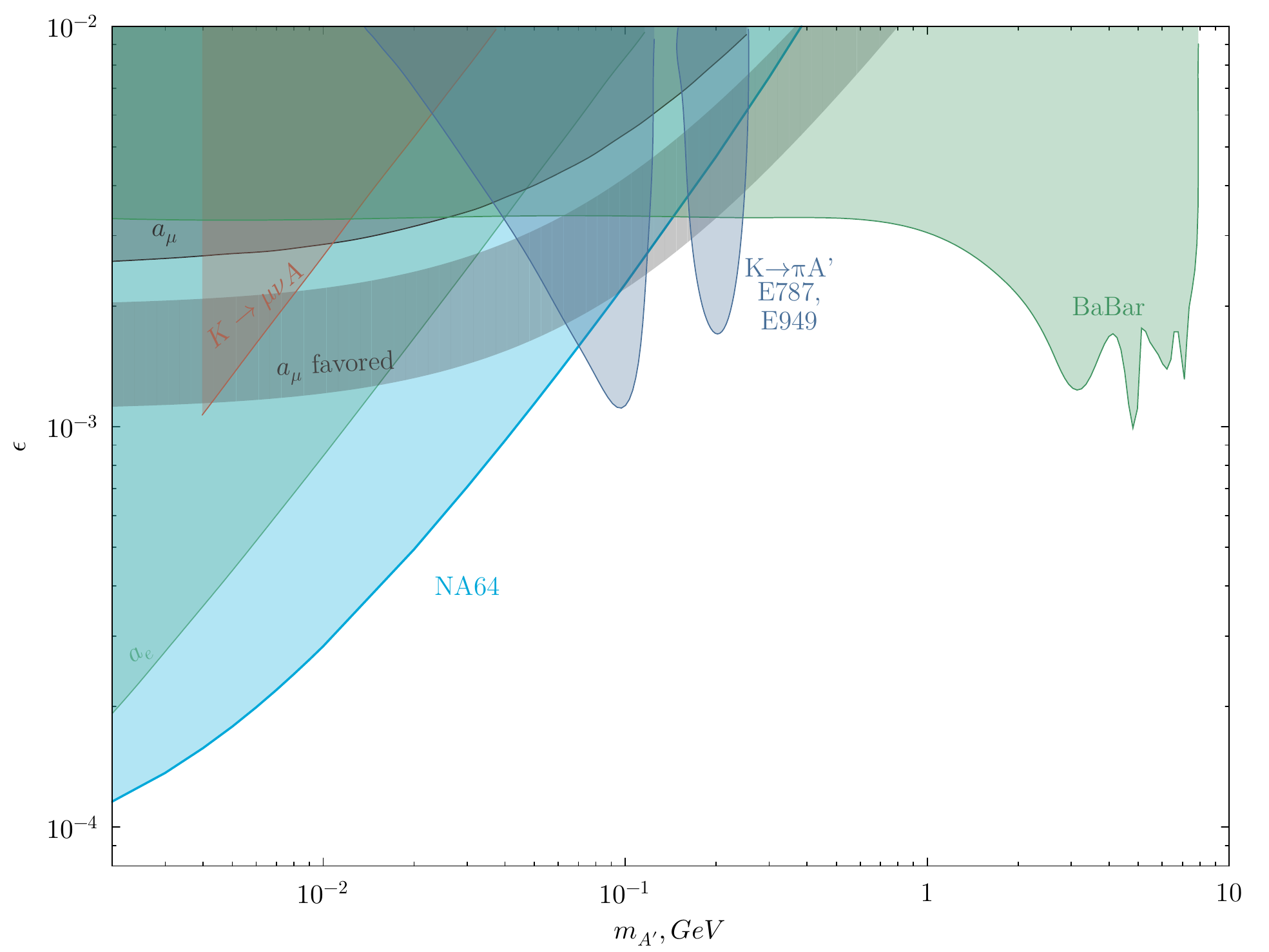}
\caption {The NA64 90 \% C.L. exclusion region in the ($m_{A'}, \epsilon$) plane.  
 Constraints from the BaBar \cite{Izaguirre:2014bca, Aubert:2008as},   and E787+ E949  experiments \cite{hd, recoll}, as well as muon  $\alpha_\mu$ favored area 
 are also shown. Here, $\alpha_\mu =\frac{g_\mu-2}{2}$.
 For more limits obtained from indirect searches and 
 planned measurements see e.g. Refs. \cite{report}.
  \label{exclinv}}
\end{center}
\end{figure} 
\par The  $m_{A'}$-dependent upper limit on the mixing $\epsilon$ is calculated as follows. 
For a given number $n_{eot}$ and the mass $m_{A'}$, the  number of  signal events $N_{A'}$ expected from the reaction \eqref{ea} in the signal box  is given by:
\begin{equation}
\Na =n_{eot} \cdot n_{A'}(\epsilon,\ma, \Delta E_{A'})\cdot \epsilon_{A'}(\ma,\Delta E_{A'}) 
\label{nev}
\end{equation} 
where $n_{A'}(\epsilon, \ma, \Delta E_{A'})$  is the yield of  $A'$s with the  coupling
$\epsilon$, mass $\ma$, and  energy in the range $\Delta E_{A'}$,  $0.5 E_0 < E_{A'} < E_0$,  per e-m shower  generated by a single 100 GeV electron in the ECAL \cite{gkkk}. These events corresponds to the missing
energy $0.5 E_0 < E_{miss} < E_0$.  The overall signal efficiency,    
$\epsilon_{A'}$ is weakly  $m_{A'},E_{A'}$ dependent and  is given by the 
product of efficiencies accounting for the NA64 geometrical acceptance (0.97), the analysis efficiency ($\simeq 0.8$) which is slightly $m_{A'}$ dependent,   veto V2 (0.96)  and HCAL signal efficiency (0.94) and the acceptance loss due to pile-up ($\simeq 8\%$ for BGO and $\simeq 7\%$ for PbSc runs).
The number of collected $n_{eot}= 2.75\cdot 10^{9}$ was estimated based on the recorded number of reference events  from the e-m $e^- Z$ interactions in the target taking into account dead time. 
 The acceptance of the signal events was evaluated by taking  all relevant momentum and angular distributions into account. 
The $A'$ yield calculated as  described in Ref.\cite{gkkk}
  was  cross-checked  with calculations of Ref.\cite{Izaguirre:2014bca}.  The $\simeq 10\%$ discrepancy between these  two calculations was accounted for as systematic uncertainty in $n_{A'}(\epsilon, \ma, \Delta E_{A'})$ due to a possible difference in treatment of the e-m shower development.  
To estimate additional uncertainty in the $A'$ yield prediction,  
the cross-check between a clean sample of $\simeq 5\cdot 10^3$ observed  and MC predicted $\mu^+ \mu^-$ events  with $E_{ECAL} \lesssim 60$ GeV  was made, resulting in  $\simeq 15\%$ difference  in  the dimuon yield.  
The number  of $A'$ and dimuon events 
are both proportional to the square of the Pb  nuclear form factor $F(q^2)$ and are sensitive to its shape. 
As the  mass $(\ma \simeq m_\mu)$ and $q^2$ $(q\simeq \ma^2 /E_{A'}\simeq m_\mu^2 /E_{\mu})$ ranges  for both reactions are similar, the observed difference can be interpreted as due to the accuracy of the  dimuon
yield calculation for  heavy nuclei and, thus can be conservatively  accounted for as  additional systematic uncertainty in $n_{A'}(\epsilon, \ma, \Delta E_{A'})$.
The V2 and HCAL signal efficiency was  defined as a fraction of events below the corresponding zero-energy thresholds. The shape of the energy distributions in these detectors  from the leak of signal shower energy from  the ECAL  was simulated  for different $A'$ masses \cite{gkkk} and cross-checked with measurements at the 
$e^- $ beam. 
The uncertainty in the V2 and HCAL efficiency for the signal events, dominated mostly by the pile-up effect  from penetrating hadrons in the high intensity PbSc run,  was estimated to be $\simeq 3\%$.
 The trigger (SRD) efficiency  is measured in unbiased random samples of events that bypass the trigger (SRD) selection and the  uncertainty is 2\% (3\%).  Other effects, e.g. 
$e^-$ loss due to conversion into $e^- \g$ pair  in the upstream detector material were measured to be $\lesssim 3\% $ (2\% uncertainty). Finally, the dominant source of  systematic errors on the expected number of signal events comes from the uncertainty in the estimate of the  yield $n_{A'}(\epsilon, \ma, \Delta E_{A'})$ (19\%). 
The overall signal  efficiency $\epsilon_{A'}$ varied from 0.69$\pm$ 0.09 to 0.55$\pm$0.07 decreasing for the 
higher  $A'$ masses. 
\par In accordance with the  $CL_s$ method \cite{90cl}, for zero observed events the 
90\% C.L. upper limit for the  number of signal events  is $N_{A'}^{90\%}(m_{A'})=2.3$. Taking this and  Eq.(\ref{nev}) into account  and using the relation
 $ N_{A'}(m_{A'}) < N_{A'}^{90\%}(m_{A'}) $  results in  the $90\%$ C.L. exclusion area in the 
($m_{A'};\epsilon $) plane shown in Fig. \ref{exclinv}. The limits are determined mostly by the number of 
accumulated eot.  
These results exclude the invisible $A'$ as an explanation of the $g_\mu-2$ muon anomaly for the masses $m_{A'} \lesssim 100$ MeV. 
Moreover, the results also allow to restrict other models with light particles interacting with electron and decaying predominantly to invisible modes. For instance for light scalar particle $s$ with 
the interaction  $ L_{es} = s\bar{e}(h_s + h_{as}i\gamma_5)e$
the bound on     $\epsilon_s$      ($\epsilon_s^2 \alpha   \equiv \frac{h^2_s + h^2_{as}}{4\pi}$) 
is approximately 1.5 times weaker than the one obtained on $\epsilon$ for the model with 
light vector bosons \cite{NA64}. Here $h_s$ and $h_{as}$  are 
scalar and pseudoscalar Yukawa coupling constants of the light scalar field $s$ with electron 
field $e$, respectively. 

\par We gratefully acknowledge the support of the CERN management and staff 
and the technical staffs of the participating institutions for their vital contributions. 
 This work was supported by the HISKP, University of Bonn (Germany), JINR (Dubna), MON and RAS (Russia), 
SNF (Switzerland), and  grants FONDECYT 1140471 and 1150792, Ring ACT1406 and Basal FB0821 CONICYT (Chile).
Part of the work on MC simulations was supported by the RSF grant 14-12-01430.  
 We thank S. Andreas and  A. Ringwald  for their contribution at the earlier stage of the project, and V.Yu. Karjavin,  J. Novy, V.I. Savrin, and I.I. Tkachev for their help. We thank COMPASS DAQ group and the Institute for Hadronic Structure and Fundamental Symmetries of TU Munich for the technical support.

\end{document}